\documentclass[preprint2]{aastex}

\begin{document}

\title{Binary Evolution Constraints from Luminosity Functions of LMXBs}

\author{R. Voss}
\affil{Max-Planck-Institut f\"ur extraterrestrische Physik,
  Giessenbachstrasse, 85748 Garching, Germany \&
  Radboud University Nijmegen, Department of Astrophysics/IMAPP, P.O. Box 9010, NL-6500 GL Nijmegen, The Netherlands}

\begin{abstract}
The formation and evolution of low-mass X-ray binaries (LMXBs) 
is not well understood. 
The properties of a population of LMXBs depend on a number of
uncertain aspects of binary evolution, and population studies 
offers a relatively new way of probing binary interactions. 
We have studied the shape of the faint end of the X-ray luminosity
function (LF) of LMXBs in nearby galaxies with Chandra and in the Milky 
Way using the Swift all-sky monitor. We find
a clear difference between the LF of LMXBs in globular clusters (GCs) 
and those outside, with a relative lack of faint GC sources.
This indicates a difference in the composition of the two populations.
\end{abstract}

%%%%%%%%%%%%%%%%%%%%%%%%%%%%%%%%%%%%%%%%%%%%
%% MAINMATTER
%%%%%%%%%%%%%%%%%%%%%%%%%%%%%%%%%%%%%%%%%%%%

\section{Introduction}
The LF of bright X-ray sources ($L_{x}>10^{35}$ erg s$^{-1}$) in old 
stellar populations is dominated by LMXBs. Single galaxies can contain
hundreds of these, and with Chandra it has become possible to study the
properties of the population statistically. The average LF has been
shown to be steep at high luminosities ($>$ few times $10^{37}$ erg s$^{-1}$)
with a (differential) power-law slope of 1.8-2.5 \citep{Gilfanov2004,Kim2004}. 
At lower luminosities
the average LF flattens to a slope of $\sim$1 \citep{Gilfanov2004}.\\
The LMXBs can be divided into two different categories,
based on their location. \textit{Field} LMXBs are found in low-density
regions of galaxies, with a spatial distribution following the density 
of stars in each galaxy. In such a low-density environment binaries evolve
without ``noticing'' other stars, and it is therefore believed that the
LMXBs have evolved from primordial binaries. The other category,
\textit{GC} LMXBs are found in GCs or similar dense stellar systems, where
the rate of stellar interactions is high. These systems are believed to
be created through encounters between two or more stars, explaining the
much higher specific density of LMXBs per unit stellar mass in this
environment.

\section{The Milky Way}
The population of LMXBs in the Milky Way was studied by \citep{Grimm2002} 
using the
RXTE All Sky Monitor (ASM). They emphasized the properties that an outside 
observer would see, which makes their study ideal for comparison with Chandra
observations of other galaxies. However, the RXTE ASM only observes photons 
in the 2-10 keV band, and Chandra and XMM-Newton are also limited to observing
photons below $\sim10$ keV. Many LMXBs emit a significant 
fraction of their X-ray luminosity in more energetic X-rays, and it is
important to understand this contribution when modelling the properties
of the population of LMXBs.

We used the Swift BAT instrument to survey the population of LMXBs
in the Galactic plane \citep{Voss-Swift}.
The Swift BAT instrument represents a major improvement in sensitivity 
for imaging of the hard X-ray sky. BAT is a coded mask, wide field of view 
telescope sensitive in the 15--200 keV energy range.
BAT's main purpose is to locate Gamma-Ray Bursts (GRBs).
While chasing new GRBs, BAT surveys the hard X-ray sky with an 
unprecedented sensitivity. Thanks to its wide FOV and its pointing strategy, 
BAT monitors continuously up to 80\% of the sky every day. Therefore
the light curves of all sources are sampled regularly in a manner
similar to the RXTE ASM. Many X-ray sources are highly variable on a 
variety of timescales, and therefore regular sampling is important for
deriving the average properties of objects, as opposed to pointed
observations that are useful for deriving the physical properties of
objects at specific times.

\begin{figure}
  \includegraphics[height=.3\textheight]{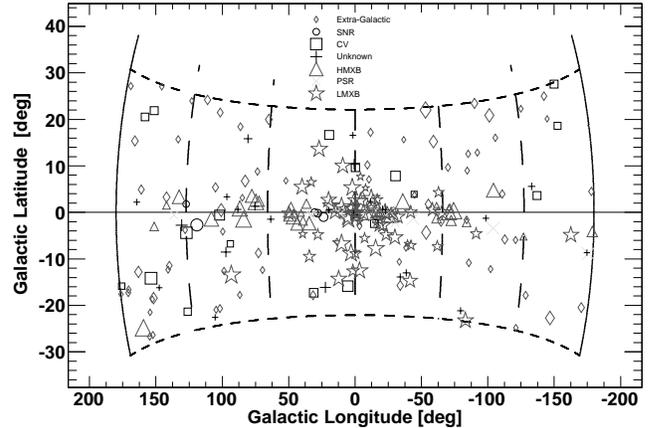}
  \caption{Identifications in the Swift BAT Galactic plane survey.}
\end{figure}

We used published identifications of the
individual sources to distinguish source types and estimate their
distances. The resulting map of sources is shown in figure 1. The
LMXBs are seen to concentrate on the bulge region, as expected for
a population following the density of stars. The sensitivity of the 
survey varies with the direction and the luminosity
of the X-ray sources. Following \citep{Grimm2002} we account for this by
setting up a model for the Galaxy, and for the range of luminosities
investigated we estimate the fraction of the Galaxy that is visible.
 For all directions we used the local background to estimate the 
limiting flux detectable by our survey, and used this to create
a sensitivity map. For a given X-ray luminosity and direction, this
enabled us to calculate the maximum distance, for which an X-ray binary
is observable. However, to identify an X-ray source as an X-ray binary, 
and to determine the distance, it is necessary to have an optically 
identified counterpart.
\citet{Grimm2002} estimated that above a distance of 10 kpc from the sun,
the optical identification of X-ray binaries becomes incomplete. We adopted
this result and limited our survey to this distance, irrespective of the
X-ray brightness of the X-ray binaries. However towards the 
galactic bulge, source confusion and extinction are serious and 
optical/IR identifications might be incomplete beyond $\sim2-3$ kpc.
 
\begin{figure}
  \includegraphics[height=.3\textheight]{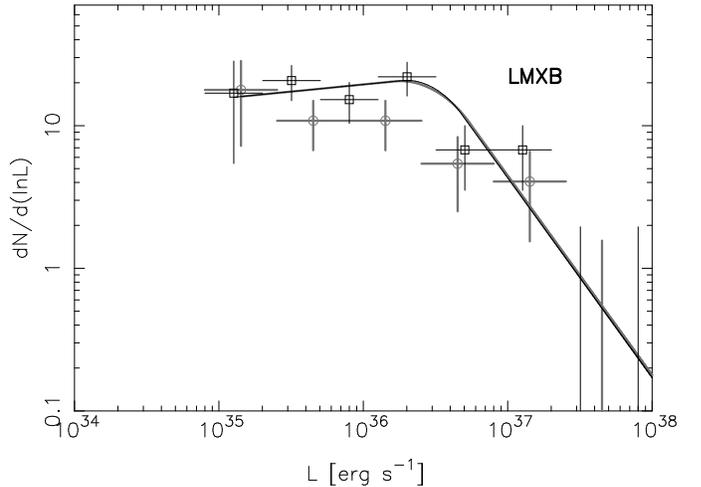}
  \caption{The luminosity function of Galactic LMXBs in the 15-55 keV band.
  Squares indicate the entire survey, whereas the inner 10 deg of the
bulge were excluded for the circles.}
\end{figure}

By combining the X-ray and optical
limits with the model of the Galaxy, we estimated the fraction of the
Galaxy observable as a function of source luminosity. 
The total Galactic luminosity function of LMXBs was then found
by correcting the observed luminosity functions for the fraction of the
Galaxy probed by our survey. The outcome is shown in 
figure 2.
Also shown in these figures are the LFs of LMXBs obtained
if the inner 10 deg of the bulge are excluded from our analysis, to assess
the effects of source confusion. The LF is somewhat different with
a lower normalization around $10^{36}$ erg s$^{-1}$. At both lower and higher
luminosities the results are in agreement with the sample including the
inner bulge. This is somewhat surprising as incompleteness due to a lack
of optical IDs is expected to lead to the opposite effect, and could indicate
a higher normalization of LMXBs per unit stellar mass in the bulge than
in the disk. However, the statistical uncertainties,
together with the uncertainties of distance determination and the mass
distribution of the Galaxy (both of which are difficult to quantify), 
are too large for such a conclusion to be significant. We note that
recent results \citep{Kim2010} indicate that the LMXB luminosity functions
are age dependent at bright end ($>10^{38}$ erg s$^{-1}$). As the bulge
and the disk have different star formation histories some variation in
the observed LFs is not unexpected.

The LF function displays roughly the same shape as the one found by
\citep{Grimm2002} in the 2-10 keV band. However, at luminosities 
above $\sim10^{37}$
erg s$^{-1}$ the 15-55 keV LF becomes much steeper, as the more
luminous states of LMXBs are soft.

\section{Andromeda and Centaurus A}
Andromeda and Centaurus A are the best available 
external environments for studying the populations of LMXBs. Both
are sufficiently nearby for LMXBs to be observed at faint luminosities
($<10^{36}$ erg s$^{-1}$) and massive enough to host significant populations.
Importantly, a large part of their mass is situated in regions with little or no
recent star formation, and the observations are therefore not contaminated
by high-mass X-ray binaries or supernova remnants. We studied the LF of the 
LMXBs in the two galaxies \citep{Voss-cena1,Voss-m31,Voss-cena2}. 
The samples were contaminated by background AGNs. 
We used the fact that the distribution
of these sources is flat on the scale of a single galaxy, and that their
$\log N - \log S$ distribution is known from deep-field surveys to
subtract their contribution statistically from our samples. We
estimated the sensitivity of the observations over the observed fields,
based on the behaviour of the varying Chandra point-spread function and 
exposure, and used this to correct our results for incompleteness. 
The resulting LFs are shown in figure 3. In this figure, the LFs are
shown separately for the field and the GC sources. A clear difference
can be seen, with a relative dearth of faint sources in the GCs.
The results are significant above the 3 $\sigma$ level. Investigations
of more distant elliptical galaxies have confirmed this difference
between the LFs of the GC and the field LMXBs \citep{Kim2009}.

\begin{figure}
  \includegraphics[height=.25\textheight]{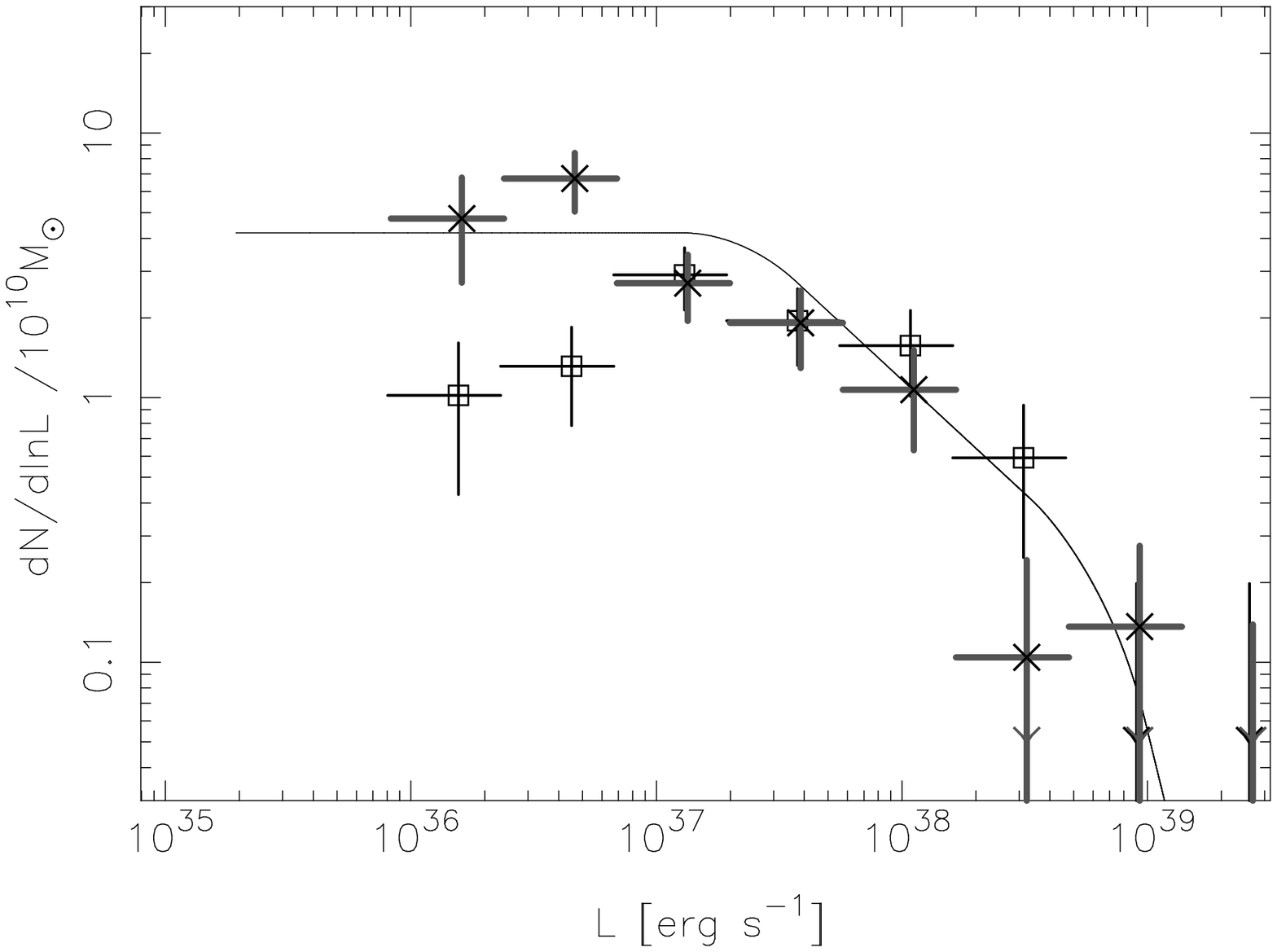}
  \includegraphics[height=.25\textheight]{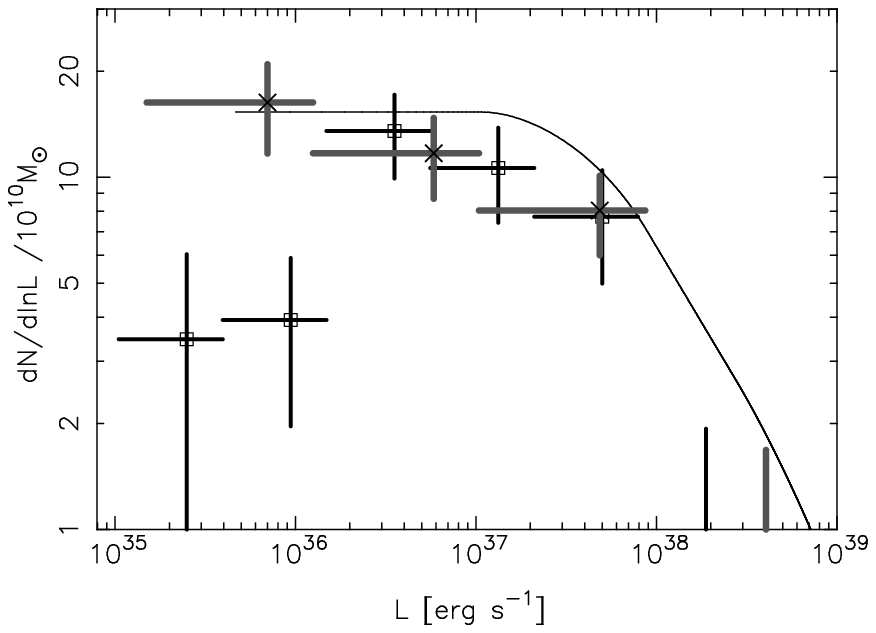}
  \caption{Luminosity functions of LMXBs in Centaurus A (left) and
  the bulge of M31 (right). The samples have been divided into
  GC LMXBs (squares) and field LMXBs (Xs).}
\end{figure}

\section{The spatial distribution of LMXBs in the bulge of M31}
The spatial distribution of LMXBs approximately follows the
density of the old stellar component in galaxies \citep{Gilfanov2004}. 
In Cen A we found good agreement with the radial profile of the LMXBs
and the stellar mass traced by the K-band light. However, 
in the inner part of the M31 bulge the radial profile of the X-ray
sources clearly deviate from the stellar density distribution 
\citep{Voss-coll}. In the central 1 arcmin there are 29 sources, 
whereas only 8.4
sources would be expected from an extrapolation of the radial
profile at larger distances from the centre. The relative excess
becomes larger at smaller radii. The profile of the sources is
shown in figure 4. Also shown is the distribution of field sources
(dashed line) extrapolated inwards from an annulus with 1-12 arcmin 
distance from the centre, and the distribution of GC sources.

\begin{figure}
  \includegraphics[height=.3\textheight,angle=270]{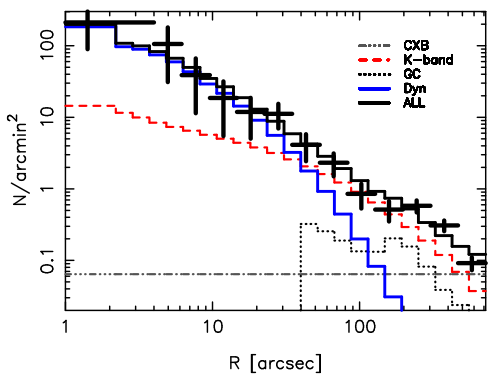}
  \caption{The spatial distribution of X-ray sources in the bulge of M31,
    with models for the contributions from various types of sources:
    background AGNs (CXB), Field LMXBs (K-band), GC LMXBs (GC) and sources
formed dynamically in the inner bulge (Dyn).}
\end{figure}

We proposed that the surplus LMXBs are formed by dynamical interactions
(similar to the formation in GCs) in the dense inner bulge of M31.
The encounter rate is approximately proportional to the integral of 
$\rho^2/\sigma_{v}$ over a stellar system, where $\rho$ is the stellar 
density and $\sigma_{v}$ is the velocity dispersion. 
The bulge is less dense than the typical GCs hosting LMXBs and
has a higher velocity dispersion (by a factor of $\sim10$), both of which
decreases the encounter rate per unit mass. However, the bulge mass is very 
large and the
rate of encounters is therefore still high enough to potentially produce
LMXBs. The expected spatial distribution of these LMXBs is indicated
by the line marked ``Dyn'' in figure 4, which provides a very good
fit to the missing component. We studied the different encounter
types which can possible lead to the formation of LMXBs to understand
if they can succesfully create LMXBs in a high velocity dispersion 
environment like the M31 bulge. The study confirmed the possibility
that the sources indeed are formed through dynamical interactions.
It was furthermore found that the LF of the surplus sources follows
the same behaviour as the LF of GC LMXBs, with a deficit of faint
sources.

\section{Discussion and conclusions}
We investigated the LF and spatial distribution of LMXBs in the
Milky Way, M31 and Cen A. These galaxies are the only massive
galaxies in which the LF of LMXBs can be studied at the faint
end ($<10^{36}$ erg s$^{-1}$). We have shown that the LF
flattens significantly at luminosities below few times $10^{37}$ erg s$^{-1}$.
We have also found that the LF of field sources is different from the
LF of the LMXBs found in GCs. This indicates that the two populations
are formed differently, and is a strong argument against the idea that
all LMXBs are formed in GCs \citep{White2002}. 
Furthermore the difference provides an
additional constraint that can be important for modelling the formation
and evolution of the LMXBs, and indicates that the bright and faint
systems are not simply different evolutionary states of the same
type of LMXBs.\\
We have also found a new population of LMXBs, formed dynamically
in the central bulge of M31. This population shows up as a surplus
of X-ray sources in the central 1 arcmin of M31, and the spatial
distribution is in agreement with our theoretical model.
The LF displays the same dearth of faint sources as the LF of
dynamically formed LMXBs in GCs.\\

\bibliographystyle{apj}
\bibliography{apj-jour,biblio}
\end{document}